\magnification \magstep1
\raggedbottom
\openup 4\jot
\voffset6truemm
\headline={\ifnum\pageno=1\hfill\else
\hfill{\it Essential self-adjointness in one-loop quantum cosmology}
\hfill \fi}
\centerline {\bf ESSENTIAL SELF-ADJOINTNESS IN}
\centerline {\bf ONE-LOOP QUANTUM COSMOLOGY}
\vskip 1cm
\centerline {Giampiero Esposito$^{1,2}$,
Hugo A. Morales-T\'ecotl$^{3}$ and Luis O. Pimentel$^{3}$}
\vskip 0.3cm
\noindent
{\it ${ }^{1}$Istituto Nazionale di Fisica Nucleare,
Sezione di Napoli, Mostra d'Oltremare Padiglione 20,
80125 Napoli, Italy;}
\vskip 0.3cm
\noindent
{\it ${ }^{2}$Dipartimento di Scienze Fisiche, Mostra d'Oltremare
Padiglione 19, 80125 Napoli, Italy;}
\vskip 0.3cm
\noindent
{\it ${ }^{3}$Universidad Aut\'onoma Metropolitana Iztapalapa,
Departamento de F\'{\i}sica, Apartado}
\vskip 0.1cm
\noindent
{\it Postal 55-534,
09340 M\'exico D. F., M\'exico.}
\vskip 0.3cm
\noindent
{\bf Abstract.} The quantization of closed cosmologies makes
it necessary to study squared Dirac operators on closed
intervals and the corresponding quantum amplitudes. This
paper shows that the proof of essential self-adjointness
of these second-order elliptic operators is related to
Weyl's limit point criterion, and to the properties of
continuous potentials which are positive near zero and are
bounded on the interval $[1,\infty[$.
\vskip 4cm
\leftline {PACS numbers: 0290, 9880}
\vskip 100cm
\leftline {\bf 1. Introduction}
\vskip 1cm
\noindent
One of the fundamental problems of quantum theory is to determine
whether the closure of the Hamiltonian operator has
self-adjoint extensions [1]. Since each extension leads to
a different physics, the problem is not just of technical
nature, but lies at the very heart of applied quantum
theory [1]. The theory of extensions of symmetric operators
is by now a rich branch of modern mathematical physics,
and its basic theorems, jointly with many examples, can be
found in [1, 2].

On the other hand, in recent years,
lots of work have been done on one-loop
quantum cosmology and on a rigorous theory of the
semiclassical amplitudes when the Dirac operator is considered
and local or spectral boundary conditions
are imposed [3--7]. Within this
framework, the Dirac operator is a first-order elliptic
operator which maps primed spinors to unprimed spinors,
and the other way around [7, 8]. Using two-component spinor
notation [8], the local boundary conditions motivated by
local supersymmetry take the form [3, 5--8]
$$
\sqrt{2} \; {_{e}n_{A}^{\; \; A'}} \; \psi^{A}
= \pm {\widetilde \psi}^{A'}
\; \; \; \; {\rm at} \; \; \; \; \partial M
\eqno (1.1)
$$
where the independent spinor fields
$\pmatrix {\psi^{A} \cr {\widetilde \psi}_{A'}\cr}$
represent a massless spin-1/2 field, and
${_{e}n_{A}^{\; \; A'}}$ is the Euclidean normal to the
boundary [3, 5]. In [3, 5] it was shown that, on imposing the
conditions (1.1) in the case of flat Euclidean backgrounds
with boundary, a first-order differential operator exists
which is symmetric and has self-adjoint extensions
(see also [7, 8]).

Spectral boundary conditions rely instead on a non-local
analysis, i.e. the separation of the spectrum of the
intrinsic three-dimensional Dirac operator
${\cal D}_{AB}={_{e}n_{AB'}} \; e_{B}^{\; \; B'j}
\; { }^{(3)}D_{j}$ of the boundary into its positive and
negative eigenvalues [4, 5].
For example, in the underlying
classical theory in the presence of a 3-sphere boundary,
the regular modes of the massless field are just the ones
which multiply harmonics having positive eigenvalues for
${\cal D}_{AB}$ on $S^{3}$. In the corresponding quantum
boundary-value problem, only half of the fermionic field can
be freely specified at the boundary (otherwise the problem
would be overdetermined), and this is given by the modes
which multiply the harmonics on $S^{3}$ having eigenvalues
${1\over 2}(n+{3\over 2})$, with $n=0,1,2,...$ [3--5].
With the notation in [4, 5] one thus writes
$$
\psi_{(+)}^{A}=0
\; \; \; \; {\rm at} \; \; \; \; S^{3}
\eqno (1.2)
$$
$$
{\widetilde \psi}_{(+)}^{A'}=0
\; \; \; \; {\rm at} \; \; \; \; S^{3}.
\eqno (1.3)
$$

A naturally occurring question is whether
a {\it unique} self-adjoint extension for the spin-1/2
boundary-value problem exists, since otherwise different
self-adjoint extensions would lead to different spectra and
hence the trace anomaly would be ill defined. We shall see
that this is not the case, and hence the boundary conditions
are enough to determine a unique, real and positive
spectrum for the squared Dirac operator (out of which the
$\zeta(0)$ value can be evaluated as in [3--5]).

For this purpose, section 2 presents a brief review of the
boundary-value problem for the massless spin-1/2 field.
Section 3 proves essential self-adjointness of the squared
Dirac operator with spectral boundary conditions,
whilst concluding remarks are presented in section 4.
\vskip 1cm
\leftline {\bf 2. The spin-1/2 problem}
\vskip 1cm
\noindent
Following [3--6], we consider flat Euclidean 4-space bounded
by a 3-sphere of radius $a$. The spin-1/2 field, represented
by a pair of independent (i.e. not related by any
conjugation) spinor fields $\psi^{A}$ and
${\widetilde \psi}^{A'}$, is expanded on a family of
3-spheres centred on the origin as [3--6]
$$
\psi^{A}={1\over 2\pi}\tau^{-{3\over 2}}\sum_{n=0}^{\infty}
\sum_{p,q=1}^{(n+1)(n+2)}\alpha_{n}^{pq}
\Bigr[m_{np}(\tau)\rho^{nqA}+{\widetilde r}_{np}(\tau)
{\overline \sigma}^{nqA}\Bigr]
\eqno (2.1)
$$
$$
{\widetilde \psi}^{A'}={1\over 2\pi}\tau^{-{3\over 2}}
\sum_{n=0}^{\infty}\sum_{p,q=1}^{(n+1)(n+2)}
\alpha_{n}^{pq}\Bigr[{\widetilde m}_{np}(\tau)
{\overline \rho}^{nqA'}+r_{np}(\tau)\sigma^{nqA'}\Bigr].
\eqno (2.2)
$$
Note that $\tau$ is the Euclidean-time coordinate which
plays the role of a radial coordinate, and the block-diagonal
matrices $\alpha_{n}^{pq}$ and the $\rho$- and
$\sigma$-harmonics are the ones described in detail in [3, 5].
If the boundary conditions (1.1) are imposed, the action
functional reduces to a purely volume term [6] and hence
takes the form [3, 5, 6]
$$
I_{E}={i\over 2}\int_{M}\biggr[{\widetilde \psi}^{A'}
\Bigr(\nabla_{AA'}\psi^{A}\Bigr)
-\Bigr(\nabla_{AA'} \; {\widetilde \psi}^{A'}\Bigr)
\psi^{A}\biggr]\sqrt{{\rm det} \; g} \; d^{4}x .
\eqno (2.3)
$$
On inserting the expansions (2.1) and (2.2) into the action
(2.3), and studying the spin-1/2 eigenvalue equations, one
finds that the modes obey the second-order differential
equations [3, 5]
$$
P_{n}{\widetilde m}_{np}=P_{n}{\widetilde m}_{n,p+1}
=P_{n}{\widetilde r}_{np}=P_{n}{\widetilde r}_{n,p+1}=0
\eqno (2.4)
$$
$$
Q_{n}r_{np}=Q_{n}r_{n,p+1}=Q_{n}m_{np}=Q_{n}m_{n,p+1}=0
\eqno (2.5)
$$
where [3, 5]
$$
P_{n} \equiv {d^{2}\over d\tau^{2}}
+\left[E_{n}^{2}-{((n+2)^{2}-{1\over 4})\over \tau^{2}}\right]
\eqno (2.6)
$$
$$
Q_{n} \equiv {d^{2}\over d\tau^{2}}
+\left[E_{n}^{2}-{((n+1)^{2}-{1\over 4})\over \tau^{2}}\right]
\eqno (2.7)
$$
and $E_{n}$ are the eigenvalues of the mode-by-mode form of the
Dirac operator [3, 5].

The modes are regular at the origin $(\tau=0)$, whilst at the
3-sphere boundary $(\tau=a)$ they obey the following conditions
(which result from (1.1)):
$$
-i \; m_{np}(a)=\epsilon \; {\widetilde m}_{n,p+1}(a)
\eqno (2.8)
$$
$$
i \; m_{n,p+1}(a)=\epsilon \; {\widetilde m}_{np}(a)
\eqno (2.9)
$$
$$
-i \; {\widetilde r}_{np}(a)=\epsilon \; r_{n,p+1}(a)
\eqno (2.10)
$$
$$
i \; {\widetilde r}_{n,p+1}(a)=\epsilon \; r_{np}(a)
\eqno (2.11)
$$
where $\epsilon \equiv \pm 1$.

In the case of spectral boundary conditions, (1.2) and
(1.3) imply instead that
$$
m_{np}(a)=0
\eqno (2.12)
$$
$$
r_{np}(a)=0 .
\eqno (2.13)
$$
Thus, one studies the one-dimensional operators $Q_{n}$
defined in (2.7), and the eigenmodes are requested to be
regular at the origin, and to obey (2.12) and (2.13) on
$S^{3}$. From now on, we will focus on the spectral case
only, since it makes it possible to use the theorems described
in the following section.
\vskip 1cm
\leftline {\bf 3. Essential self-adjointness}
\vskip 1cm
\noindent
The previous section shows that we have to
study, for all $n \geq 0$, the differential operators
$$
{\widetilde Q}_{n} \equiv -{d^{2}\over d\tau^{2}}
+{((n+1)^{2}-{1\over 4})\over \tau^{2}}.
\eqno (3.1)
$$
These are particular cases of a large class of operators
considered in the literature. They can be studied by using
the following definitions and theorems from [1, 2]:
\vskip 0.3cm
\noindent
{\bf Definition 3.1} The function $V$ is in the
{\it limit circle} case at zero if for some, and therefore
all $\lambda$, {\it all} solutions of the equation
$$
-\varphi''(x)+V(x)\varphi(x)=\lambda \varphi(x)
\eqno (3.2)
$$
are square integrable at zero.
\vskip 0.3cm
\noindent
{\bf Definition 3.2} If $V(x)$ is not in the limit circle
case at zero, it is said to be in the {\it limit point}
case at zero.
\vskip 0.3cm
\noindent
{\bf Theorem 3.1} (Weyl's limit point - limit circle
criterion) Let $V$ be a continuous real-valued function on
$(0,\infty)$. Then ${\cal O} \equiv -{d^{2}\over dx^{2}}
+V(x)$ is essentially self-adjoint on $C_{0}^{\infty}(0,\infty)$
if and only if $V(x)$ is in the limit point case at both
zero and infinity.
\vskip 0.3cm
\noindent
{\bf Theorem 3.2} Let $V$ be continuous and {\it positive}
near zero. If $V(x) \geq {3\over 4}x^{-2}$ near zero, then
$\cal O$ is in the limit point case at zero.
\vskip 0.3cm
\noindent
{\bf Theorem 3.3} Let $V$ be differentiable on $]0,\infty[$
and bounded above by $K$ on $[1,\infty[$. Suppose that
\vskip 0.1cm
\noindent
(i) $\int_{1}^{\infty}{dx \over \sqrt{K-V(x)}}=\infty$.
\vskip 0.1cm
\noindent
(ii) $V'(x) {\mid V(x) \mid}^{-{3\over 2}}$ is bounded near
infinity.
\vskip 0.1cm
\noindent
Then $V(x)$ is in the limit point case at $\infty$.

In other words, a necessary and sufficient condition for the
existence of a unique self-adjoint extension, is that the
eigenfunctions of $\cal O$ should fail to be square integrable
at zero and at infinity. We will not give many technical
details, since we are more interested in the applications of
the general theory. However, we find it helpful for the
general reader to point out that,
when $V(x)$ takes the form ${c\over x^{2}}$ for $c>0$,
theorem 3.2 can be proved as follows [1]. The equation
$$
-\varphi''(x)+{c\over x^{2}}\varphi(x)=0
\eqno (3.3)
$$
admits solutions of the form $x^{\alpha}$, where
$\alpha$ takes the values
$$
\alpha_{1}={1\over 2}+{1\over 2}\sqrt{1+4c}
\eqno (3.4)
$$
$$
\alpha_{2}={1\over 2}-{1\over 2}\sqrt{1+4c}.
\eqno (3.5)
$$
When $\alpha=\alpha_{1}$ the solution is obviously
square integrable at zero. However, when
$\alpha=\alpha_{2}$, the solution is square integrable
at zero if and only if $\alpha_{2}>-{1\over 2}$,
which implies $c<{3\over 4}$. By virtue of definitions
3.1 and 3.2, this means that $V(x)$ is in the limit point
at zero if and only if $c \geq {3\over 4}$.

However, in our problem we are interested in the closed
interval $[0,a]$, where $a$ is finite. Thus, the previous
theorems can only be used after relating the original
problem to another one involving the infinite interval
$(0,\infty)$. For this purpose, we perform a
double change of variables. The first one affects the
independent variable, whereas the second one is performed
on the dependent variable so that the operator takes the form
(3.2) with the appropriate (semi)infinite domain.

Given (3.3) with $x\in [0,a]$ (see (3.1)) we perform the change
$$
x \equiv a\left(1-e^{-y}\right)\,\Rightarrow y\in (0,\infty).
\eqno (3.6)
$$
This yields the equation
$$
\ddot{\varphi}(y) + \dot{\varphi}(y) -
{c(n) \over (e^{y}-1)^{2}} \varphi(y)
=0 \; \; \; \; y\in (0,\infty)
\eqno (3.7)
$$
where $c(n) \equiv (n+1)^{2}-{1\over 4}$ and the
{\it dot} denotes differentiation with respect to $y$.
The further change
$$
\varphi(y) \equiv e^{-{y\over 2}} \chi(y)
\eqno (3.8)
$$
leads to
$$
-\ddot{\chi}(y)+ \left[{1\over 4}
+ {c(n)\over (e^{y} -1)^{2}}
\right] \chi=0
\; \; \; \; y \in (0,\infty)
\eqno (3.9)
$$
which has the structure of the left hand side of
(3.2) with a suitable domain,
so that Weyl's theorem is directly applicable.
Indeed, when $y \rightarrow \infty$, the conditions (i)
and (ii) of theorem 3.3 are clearly satisfied by the potential
in (3.9), and hence such a $V(y)$ is in the limit point
at $\infty$ (in that case, the constant of theorem 3.3 is
$K(n) \equiv {1\over 4}+{c(n)\over (e-1)^{2}}$).
Moreover, when $y \rightarrow 0$, $V(y)$ tends to
${1\over 4}+{c(n)\over y^{2}} \geq {1\over 4}
+{3\over 4}{1\over y^{2}}$, since $n \geq 0$ in (3.1).
Thus, the condition of theorem 3.2 is satisfied and we are
in the limit point at zero as well. By virtue of theorem 3.1,
these properties imply that our operators (3.1) are related,
through (3.6) and (3.8), to the second-order elliptic
operators
$$
T_{n} \equiv -{d^{2}\over dy^{2}}+{1\over 4}
+{((n+1)^{2}-{1\over 4})\over (e^{y}-1)^{2}}
\eqno (3.10)
$$
which are essentially self-adjoint on the space of
$C^{\infty}$ functions on $(0,\infty)$ with compact support
(see section 4).
\vskip 1cm
\leftline {\bf 4. Concluding remarks}
\vskip 1cm
\noindent
There is indeed a rich mathematical literature on
self-adjointness properties of the Dirac operator
(see, for example, [9] and references therein). However,
one-loop quantum cosmology needs a rigorous proof of essential
self-adjointness of {\it elliptic} operators with local or non-local
boundary conditions. Our paper represents the first step in this
direction, in the case of the squared Dirac operator with the
spectral boundary conditions (1.2) and (1.3) [4, 5].
Note that such non-local boundary conditions play a crucial
role in section 3. As shown on page 153 of [1], to prove
theorem 3.1 one has to choose at some stage a point
$c \; \in \; (0,\infty)$ and then consider the operator
$$
A \equiv -{d^{2}\over dx^{2}}+V(x)
$$
on the domain
$$
D(A) \equiv \left \{ \omega: \omega \; \in \; C^{\infty}(0,c),
\omega=0 \; {\rm near} \; {\rm zero},
\omega(c)=0 \right \}.
$$
As far as we can see, this scheme is only compatible with the
boundary conditions (2.12) and (2.13), which result from the
spectral choice (1.2) and (1.3). Thus, we should stress that
the extension to the boundary conditions (2.8)--(2.11)
resulting from (1.1) remains the main open problem in our
investigation.

In this respect, we should also emphasize that, in the
{\it differential equations} viewpoint, one focuses on the
boundary conditions, whilst in the {\it functional analysis}
viewpoint the key elements are the domains of differential
operators. Strictly, the domain corresponding to a given
choice of boundary conditions may also contain some less
well behaved functions (we thank Bernard Kay for clarifying
this point).

Moreover, it remains to be seen how to apply similar techniques to
the analysis of higher-spin fields. These are gauge fields and
gravitation, which obey a complicated set of mixed boundary
conditions [10--13]. If this investigation could be completed,
it would put on solid ground the current work on trace anomalies
and one-loop divergences
on manifolds with boundary [3--5, 10--13], and it would add evidence
in favour of quantum cosmology being at the very heart of many
exciting developments in quantum field theory, analysis and
differential geometry [3--5, 12, 13].
\vskip 1cm
\leftline {\bf Acknowledgments}
\vskip 1cm
\noindent
We are indebted to Bernard Kay for correspondence.
The first author is grateful to the European Union for
partial support under the Human Capital and Mobility
Programme. HAMT and LOP acknowledge the partial
support from CONACyT grants No. 3544--E9311
and 1861--E9212.
\vskip 1cm
\leftline {\bf References}
\vskip 1cm
\item {[1]}
Reed M and Simon B 1975 {\it Methods of Modern Mathematical Physics,
Vol. II: Fourier Analysis and Self-Adjointness} (New York:
Academic)
\item {[2]}
Weidmann J 1980 {\it Linear Operators in Hilbert Spaces}
(Berlin: Springer)
\item {[3]}
D'Eath P D and Esposito G 1991 {\it Phys. Rev.} D {\bf 43} 3234
\item {[4]}
D'Eath P D and Esposito G 1991 {\it Phys. Rev.}
D {\bf 44} 1713
\item {[5]}
Esposito G 1994 {\it Quantum Gravity, Quantum Cosmology and
Lorentzian Geometries} ({\it Lecture Notes in Physics}
{\bf m12}) (Berlin: Springer)
\item {[6]}
Esposito G, Morales-T\'ecotl H A and Pollifrone G 1994
{\it Found. Phys. Lett.} {\bf 7} 303
\item {[7]}
Esposito G {\it Dirac Operator and Eigenvalues in Riemannian
Geometry} (SISSA lectures, GR-QC 9507046)
\item {[8]}
Esposito G 1995 {\it Complex General Relativity}
({\it Fundamental Theories of Physics} {\bf 69})
(Dordrecht: Kluwer)
\item {[9]}
Chernoff P R 1977 {\it Pacific J. Math.} {\bf 72} 361
\item {[10]}
Esposito G, Kamenshchik A Yu, Mishakov I V and Pollifrone G
1995 {\it Phys. Rev.} D {\bf 52} 3457
\item {[11]}
Esposito G and Kamenshchik A Yu {\it Mixed Boundary Conditions
in Euclidean Quantum Gravity} (DSF preprint 95/23,
GR-QC 9506092, to appear in {\it Class. Quantum Grav.})
\item {[12]}
Esposito G {\it The Impact of Quantum Cosmology on Quantum
Field Theory} to appear in Proceedings of the 1995 Moscow
Quantum Gravity Seminar (DSF preprint 95/27, GR-QC 9506046)
\item {[13]}
Esposito G {\it Non-Local Properties in Euclidean Quantum
Gravity} to appear in Proceedings of the Third Workshop on
Quantum Field Theory Under the Influence of External
Conditions, Leipzig, September 1995 (DSF preprint 95/37,
GR-QC 9508056)
\vskip 100cm
\leftline {\bf Corrigendum}
\vskip 0.3cm
\leftline {2000 {\it Class. Quantum Grav.} {\bf 17} 3091}
\vskip 0.3cm
\noindent
In our paper we study the differential operator
$$
{\widetilde Q}_{n} \equiv -{d^{2}\over d\tau^{2}}
+{((n+1)^{2}-{1\over 4})\over \tau^{2}}
$$
where $\tau \in [0,a]$ and $n=0,1,2,..$, with domain given by the
functions $u$ in $AC^{2}[0,a]$ such that $u(a)=0$. The operator
${\widetilde Q}_{n}$ obeys the limit point condition at $\tau=0$,
as we prove in the paper, and the limit circle condition at
$\tau=a$. These properties, jointly with the homogeneous Dirichlet
condition at $\tau=a$, are sufficient to obtain a self-adjoint 
boundary-value problem. Thus, there is no need to change independent and 
dependent variable as we did. Moreover, our changes of variable
lead actually to the operator
$$
{\widetilde T}_{n} \equiv
a^{-2}{\rm e}^{2y}\left[-{d^{2}\over dy^{2}}+{1\over 4}
+{((n+1)^{2}-{1\over 4})\over ({\rm e}^{y}-1)^{2}}\right]
$$
where $y$ ranges from $0$ through $\infty$. The Weyl limit point-limit
circle criterion stated in our {\it Theorem 3.1} ensures that the operator
${\cal O} \equiv -{d^{2}\over dx^{2}}+V(x)$ is essentially self-adjoint on
$C_{0}^{\infty}(0,\infty)$ if and only if $V(x)$ is in the limit point at
both zero and infinity. To check this, it is indeed sufficient to study
the eigenvalue equation ${\cal O}\varphi(x)=\lambda \varphi(x)$ for a particular
value of $\lambda$, e.g. $\lambda=0$. We had instead considered the 
zero-eigenvalue equation in the transformed variables leading to
${\widetilde T}_{n}$, because ${\widetilde Q}_{n}$ is not studied on
$(0,\infty)$. But this was not enough, because the corresponding scalar 
product is no longer
$$
(u,v) \equiv \int_{0}^{1}u^{*}(\tau)v(\tau)d\tau
$$
but
$$
(u,v) \equiv \int_{0}^{\infty}u^{*}(y)v(y){\rm e}^{-2y}dy
$$
and the counterpart (if any) of the Weyl criterion for the operator
${\widetilde T}_{n}$ remains to be elaborated.

We are indebted to Stephen Fulling for having brought to our
attention our mistake and for detailed correspondence
about this problem.

\bye